\documentclass[aps,prl,superscriptaddress,amsmath,amssymb,floatfix,twocolumn]{revtex4-1}
\usepackage{graphicx}
\usepackage{subfigure}
\usepackage{color}
\usepackage{epstopdf}
\usepackage{ulem}

\begin{document}

\title{Electron correlations and superconductivity
in La$_3$Ni$_2$O$_7$ under pressure tuning}

\author{Zhiguang Liao}
\thanks{These authors contributed equally to this study.}
\affiliation{Department of Physics and Beijing Key Laboratory of Opto-electronic Functional Materials \& Micro-nano Devices,
Renmin University of China, Beijing 100872, China}

\author{Lei Chen}
\thanks{These authors contributed equally to this study.}
\affiliation{Department of Physics \& Astronomy,
Rice Center for Quantum Materials,
Rice University, Houston, Texas 77005,USA}

\author{Guijing Duan}
\thanks{These authors contributed equally to this study.}
\affiliation{Department of Physics and Beijing Key Laboratory of Opto-electronic Functional Materials \& Micro-nano Devices,
Renmin University of China, Beijing 100872, China}

\author{Yiming Wang}
\affiliation{Department of Physics \& Astronomy,
Rice Center for Quantum Materials,
Rice University, Houston, Texas 77005,USA}

\author{Changle Liu}
\affiliation{School of Engineering, Dali University, Dali, Yunnan 671003,
China}
\affiliation{Shenzhen Institute for Quantum Science and Technology and
Department of Physics, Southern University of Science and Technology,
Shenzhen 518055, China}

\author{Rong Yu}
\email{rong.yu@ruc.edu.cn}
\affiliation{Department of Physics and Beijing Key Laboratory of Opto-electronic Functional Materials \&
Micro-nano Devices, Renmin University of China, Beijing 100872, China}
\affiliation{Key Laboratory of Quantum State Construction and Manipulation (Ministry of Education),
Renmin University of China, Beijing, 100872, China}

\author{Qimiao Si}
\email{qmsi@rice.edu}
\affiliation{Department of Physics \& Astronomy,
Rice Center for Quantum Materials,
Rice University, Houston, Texas 77005,USA}

\begin{abstract}
Motivated by the recent discovery of superconductivity in La$_3$Ni$_2$O$_7$ under pressure,
we discuss the basic ingredients of a model that captures its microscopic physics under
pressure tuning. We anchor our description in terms of the spectroscopic  evidence of strong correlations in this system. In a bilayer Hubbard model including the Ni $3d$ $x^2-y^2$ and $z^2$ orbitals, we show the ground state of the model crosses over from a low-spin $S=1/2$ state to a high-spin $S=3/2$ state.
In the high-spin state, the two $x^2-y^2$ and the bonding $z^2$ orbitals are all close to half-filling,
which promotes a strong orbital selectivity in a broad crossover regime of the phase diagram
pertinent to the system. Based on these results, we construct an effective multiorbital $t$-$J$ model to describe the superconductivity of the system, and find the leading pairing channel to be an intraorbital spin singlet with a competition between the extended $s$-wave and $d_{x^2-y^2}$ symmetries. Our results highlight the role of strong multiorbital correlation effects in driving the superconductivity of La$_3$Ni$_2$O$_7$.
\end{abstract}

\maketitle


{\it Introduction.~}
The discovery of iron-based superconductors more than a decade ago provided hope
for high temperature superconductivity in a variety of transition-metal-based
materials~\cite{Kamihara_JACS_2008,Johnston_2010,Si-Hussey_2023}.
The recent discovery of superconductivity in a bilayer Ni-based compound
La$_3$Ni$_2$O$_7$, with a transition temperature of about $80$ K
 when the applied pressure exceeds $14$ GPa~\cite{SunWang_Nature_2023},
 was soon confirmed~\cite{Cheng_arXiv_2023} and
 zero resistivity
 was recently obtained~\cite{Yuan_arXiv_2023}.
 Unlike the infinite-layer nickelate
 (Sr,Nd)NiO$_2$ thin films~\cite{Li_Nature_2019}, which was expected to resemble the
  physics of the
cuprates given the valence count of Ni$^{1+}$ with a $d^{9}$ electron configuration,
a simple valence count gives Ni$^{2.5+}$ in the bilayer compound La$_3$Ni$_2$O$_7$,
corresponding to $d^{7.5}$.
These results have naturally attracted extensive interest~\cite{Yao_arXiv_2023, Hu_arXiv_2023, Werner_arXiv_2023, WangQ_arXiv_2023, Lechermann_arXiv_2023, ZhangG_arXiv_2023, Leonov_arXiv_2023, Dagotto_arXiv_2023, Kuroki_arXiv_2023, Zhang_arXiv_2023, Lu_arXiv_2023, YangF_arXiv_2023, Yang_arXiv_2023, Wang_arXiv_2023, WuYang_arXiv_2023}.

\begin{figure}[t!]
\centering\includegraphics[
width=85mm 
]{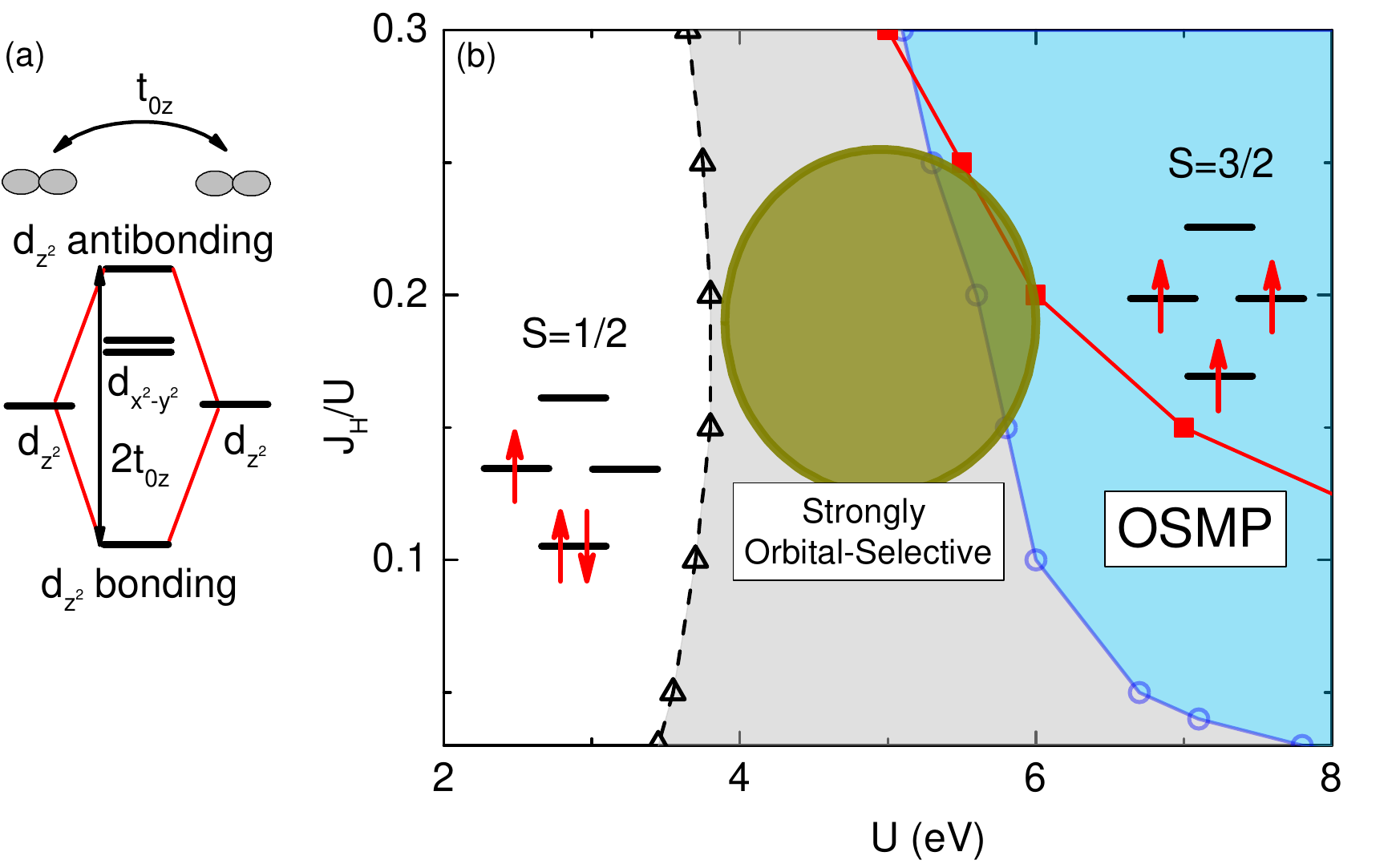}
\caption{(Color online)
(a): Sketch of formation of the bonding-antibonding MO states between the Ni $z^2$ orbitals in the top and bottom layers.
(b): Ground-state phase diagram in the $U$-$J_{\rm{H}}$ plane of the bilayer two-orbital Hubbard model for La$_3$Ni$_2$O$_7$,
calculated by the $U(1)$ slave-spin theory in the MO basis. Red line with squares shows the transition
between the low-spin $S=1/2$ state to the high-spin $S=3/2$ state in the atomic limit.
Blue regime is the orbital-selective Mott phase (OSMP) in which the $x^2-y^2$ orbitals
are Mott localized whereas the $z^2$ orbitals are itinerant (see Fig.~\ref{fig:2}(a)).
The dahsed line with triangles characterizes the low-spin to high-spin crossover,
and in the gray regime the system exhibits strong orbital selectivity (see text).
Highlighted by the golden shading in the plot is the parameter region pertinent to La$_3$Ni$_2$O$_7$.
}
\label{fig:1}
\end{figure}

Some of the key questions concern the roles of multiple orbitals and electron correlations~\cite{Wen_arXiv_2023}
for both the normal state and superconductivity of  La$_3$Ni$_2$O$_7$.
The first-principles density functional theory (DFT)
calculation~\cite{SunWang_Nature_2023,SM}
shows that the bands
near the Fermi level have mainly Ni $e_g$ orbital characters; bands
with $t_{2g}$ orbital characters are located about $2$ eV below the Fermi level.
It also reveals a strong inter-layer hopping between the two Ni $z^2$ orbitals through the apical oxygen ion,
which leads to the
formation of bonding-antibonding molecular orbitals (MOs) as illustrated in Fig.~\ref{fig:1}(a).
Considering DFT results and the simple valence count, a na\"{\i}ve picture of the electron state is as follows:
Within a two-Ni unit cell, the $t_{2g}$ orbitals are almost fully occupied, and the four $e_g$ molecular orbitals
are occupied by three electrons.
To make progress, it is important to understand the exact ground-state configuration and the low-energy
electronic degrees of freedom that underlie the normal state and
are
responsible for the superconductivity.

In this Letter, we address these important issues by studying the electron correlations
in a bilayer two-orbital Hubbard model for La$_3$Ni$_2$O$_7$.
Importantly,
we anchor our description in terms of the spectroscopically derived experimental evidence for
strong correlations~\cite{Wen_arXiv_2023}. One of our key findings is to
show that the electron correlations drive the ground state from a low-spin $S=1/2$ configuration
to a high-spin $S=3/2$ one. In the broad crossover regime of the phase diagram, the system exhibits
strong orbital
selectivity.
This regime falls in the parameter range pertinent to La$_3$Ni$_2$O$_7$,
as highlighted in  Fig.~\ref{fig:1}(b).
We clarify the nature of this crossover regime by showing that further increasing the interaction strength stabilizes an orbital-selective Mott phase (OSMP)
in which the two $x^2-y^2$ orbitals are Mott localized whereas the $z^2$ orbitals remain itinerant:
Thus, a {\it proximity} to the OSMP underlies the strong orbital selectivity seen in the crossover regime.
Our results provide the natural understanding of the
recent optical conductivity
experiments,
which provided evidence that
the electrons' kinetic energy
is less than $0.1$ of
 its noninteracting counterpart~\cite{Wen_arXiv_2023}
 and that the Drude peak contains two components~\cite{Wen_arXiv_2023}.
 Our results on the correlation effect, in turn, allow us to advance the low-energy physics that drives the superconductivity.

{\it Model and method.~}
We consider
 a bilayer multiorbital Hubbard model
for the two $e_g$ orbitals of Ni:
$ H = H_{\rm{TB}} + H_{\rm{int}}$.
Here, $H_{\rm{TB}}$ is a tight-binding Hamiltonian. 
\begin{eqnarray}
 \label{Eq:Ham_0}
  H_{\rm{TB}} && =\frac{1}{2}\sum_{i\delta_{n(z)}\alpha\beta\sigma} t^{\alpha\beta}_{\delta_{n(z)}}
 d^\dagger_{i\alpha\sigma} d_{i+\delta_{n(z)}\beta\sigma} \nonumber \\
 && + \sum_{i\alpha\sigma} (\epsilon_\alpha-\mu)
 d^\dagger_{i\alpha\sigma} d_{i\alpha\sigma},
\end{eqnarray}
where $d^\dagger_{i\alpha\sigma}$ creates an electron in orbital $\alpha$ ($\alpha=x,z$
denoting the two $e_g$ orbitals, $x^2-y^2$ and $z^2$, respectively)
with spin $\sigma$ at site $i$ of a bilayer square lattice,
$\delta_{n(z)}$ denotes the $n$-th neighboring site in the same (opposite) layer,
$\epsilon_\alpha$ refers to the energy level associated with the crystal field splittings,
and $\mu$ is the chemical potential.
The tight-binding parameters $t^{\alpha\beta}_{\delta_{n(z)}}$ and $\epsilon_\alpha$
are obtained by fitting the calculated DFT band structure and projecting to the two-$e_g$-orbital basis
in a unit cell including two Ni sites as described in the Supplemental Material (SM)~\cite{SM}.
We adjust the chemical potential so that the total electron density is $3$ per unit cell
to reflect the valence count of Ni$^{2.5+}$.
The on-site interaction $H_{\rm{int}}$ reads
\begin{eqnarray}
 \label{Eq:Ham_int} H_{\rm{int}} &=& \frac{U}{2} \sum_{i,\alpha,\sigma}n_{i\alpha\sigma}n_{i\alpha\bar{\sigma}}\nonumber\\
 &+&\sum_{i,\alpha<\beta,\sigma} \left\{ U^\prime n_{i\alpha\sigma} n_{i\beta\bar{\sigma}}\right.
 + (U^\prime-J_{\rm{H}}) n_{i\alpha\sigma} n_{i\beta\sigma}\nonumber\\
&-&\left.J_{\rm{H}}(d^\dagger_{i\alpha\sigma}d_{i\alpha\bar{\sigma}} d^\dagger_{i\beta\bar{\sigma}}d_{i\beta\sigma}
 +d^\dagger_{i\alpha\sigma}d^\dagger_{i\alpha\bar{\sigma}}
 d_{i\beta\sigma}d_{i\beta\bar{\sigma}}) \right\},~
\end{eqnarray}
where $n_{i\alpha\sigma}=d^\dagger_{i\alpha\sigma} d_{i\alpha\sigma}$.
Here,
$U$, $U^\prime$, and $J_{\rm{H}}$, respectively denote the intra-
and inter- orbital repulsion and the Hund's rule coupling, and
$U^\prime=U-2J_{\rm{H}}$
is taken~\cite{Castellani_PRB_1978}.

As already mentioned, the strong hopping between the two Ni $z^2$ orbitals in the upper
and lower layers causes bonding-antibonding MO states.
To examine this bonding effect,
we perform a transformation from the atomic orbital basis
to the bonding
MO basis, namely by defining the MO as
\begin{eqnarray}
 d_{i\alpha\sigma}^{b(a)} = \frac{1}{\sqrt{2}} (d_{i\alpha\sigma} \pm d_{i+\delta_{0z}\alpha\sigma}),
\end{eqnarray}
where the index $b(a)$ corresponds to the bonding (antibonding) MO,
and $i$ ($i+\delta_{0z}$) on the right hand side refers to a site in the top (bottom) layer.
Note that we define MOs for both $z^2$ and $x^2-y^2$ orbitals for convenience,
though the $x^2-y^2$ orbital is expected to
be largely non-bonding given the small inter-layer hopping amplitude associated with this orbital~\cite{SM}.
We can rewrite the tight-binding and interaction Hamiltonians of Eqns.~\eqref{Eq:Ham_0} and \eqref{Eq:Ham_int}
in the MO basis.
In particular,
\begin{eqnarray}
 \label{Eq:Ham_int_MO} H_{\rm{int}} &=& H_{\rm{int}}^{b-b} + H_{\rm{int}}^{b-a},
\end{eqnarray}
where $H_{\rm{int}}^{b-b}$ refers to the interactions between bonding (or antibonding) states,
whereas $H_{\rm{int}}^{b-a}$ refers to the interactions mixing the bonding and antibonding states.
The exact forms of $H_{\rm{int}}^{b-b}$ and $H_{\rm{int}}^{b-a}$ are presented in the SM~\cite{SM}.

The correlation effects of the above model in the MO basis are then investigated by using
a $U(1)$ slave-spin theory~\cite{Yu_PRB_2012, Yu_PRB_2017}. In this
approach, the $d$-electron operators are
rewritten as
$d^\dagger_{i\alpha\sigma} = S^+_{i\alpha\sigma} f^\dagger_{i\alpha\sigma}$ (here we absorb the MO index in $\alpha$),
where $S^+_{i\alpha\sigma}$ ($f^\dagger_{i\alpha\sigma}$) is a quantum $S=1/2$ spin (fermionic spinon)
operator introduced to carry the
electron's charge (spin) degree of freedom,
and $S^z_{i\alpha\sigma} = f^\dagger_{i\alpha\sigma} f_{i\alpha\sigma} - \frac{1}{2}$ is a local constraint.
At the saddle-point level, we employ a Lagrange multiplier $\lambda_{\alpha}$
to handle the constraint, and decompose the slave-spin and spinon operators.
In this way, the model rewritten in the slave-spin representation is solved by determining $\lambda_{\alpha}$
and the quasiparticle spectral weight $Z_\alpha\propto |\langle S^+_{\alpha} \rangle|^2$
self-consistently~\cite{Yu_PRB_2012, Yu_PRB_2017}.

{\it Low-spin to high-spin crossover and orbital-selective Mott physics.~}
We first diagonalize the interaction Hamiltonian of Eqn.~\eqref{Eq:Ham_int_MO} in the MO basis.
The ground state is either a low-spin $S=1/2$ state at small $J_{\rm{H}}$
or a high-spin $S=3/2$ state at large $J_{\rm{H}}$~\cite{SM}.
In the low-spin state, the $z^2$ bonding state is largely doubly occupied,
whereas the antibonding state is almost empty. The $x^2-y^2$ orbitals are near quarter filling.
In the high-spin state, on the other hand, the bonding $z^2$ and $x^2-y^2$ orbitals are all half filled,
and the antibonding $z^2$ state keeps empty. Either the low-spin or the high-spin configuration is four-fold degenerate.
For the low-spin state, the additional degeneracy comes from the doubly degenerate $x^2-y^2$ orbitals,
which can be described by an isospin $\tau=1/2$. The transition from the low-spin to high-spin state
is shown as the red line with square symbols in Fig.~\ref{fig:1}(b).
In the presence of electron hopping, this transition turns to a crossover.

\begin{figure}[t!]
\centering\includegraphics[
width=85mm 
]{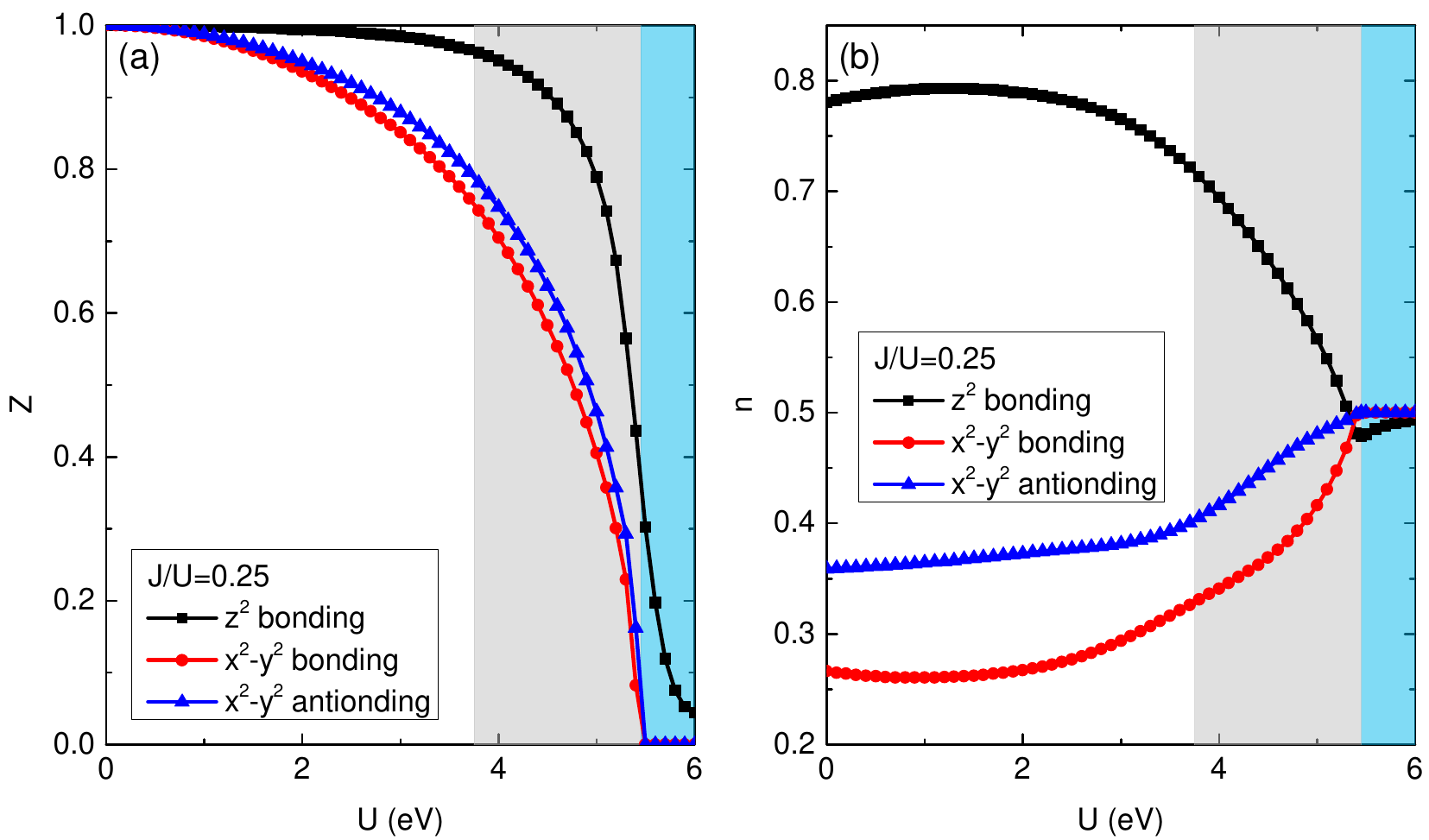}
\caption{(Color online)
Evolution of the orbital-resolved quasiparticle spectral weight $Z$
[in (a)]
and electron density $n$
[in (b)]
with increasing $U$ at $J_{\rm{H}}/U=0.25$ of the
bilayer
two-orbital model in the MO basis calculated by using the $U(1)$ slave-spin theory.
The blue and gray regimes correspond to OSMP and
metallic state with strong orbital selectivity, respectively.
}
\label{fig:2}
\end{figure}

To examine how this crossover takes place in the multiorbital Hubbard model,
we perform slave-spin calculation in the MO basis. Note that the antibonding $z^2$ state has a very low electron density $n<0.1$
at $U=0$. Accordingly, we expect it to be only weakly affected by interactions. To simplify the calculation,
we set $Z=1$ in this MO and turn off interaction terms associated with this orbital.
The results of the slave-spin calculation is summarized in the phase diagram of Fig.~\ref{fig:1}(b).
The dashed line characterizes the low-spin to high-spin crossover~\cite{SM}.
Across this crossover line with increasing $U$ to the gray regime, the system exhibits strong orbital selectivity:
As shown in Fig.~\ref{fig:2}(a), quasiparticle spectral weights and electron densities in all orbitals change drastically
and $Z_{x^2-y^2}\ll Z_{z^2}$ in this regime. Further increasing $U$, the system undergoes an orbital-selective Mott transition (OSMT)
to an OSMP at the blue line with circles. As shown in Fig.~\ref{fig:2},
in the OSMP $Z_{x^2-y^2}=0$ and $n_{x^2-y^2}=1/2$, but $Z_{z^2(b)}>0$:
The electrons in the
$x^2-y^2$ orbitals are Mott localized whereas
those
in the $z^2$ bonding states are still itinerant, though very close to the Mott localization.

The OSMP is associated with the high-spin state. One sees from Fig.~\ref{fig:2}(b) that in this state the electron densities
of the $x^2-y^2$ and $z^2$ bonding orbitals are all close to $1/2$.
If the antibonding state were to
be
completely empty, the system consisting of the other three orbitals would be exactly at half-filling and becomes Mott insulating
when the red line of transition is approached. However, at finite $U$ an OSMP is more favorable because
keeping the $z^2$ bonding orbital itinerant reduces the kinetic energy. This naturally explains why the OSMT line
almost traces the red transition line to the high-spin Mott insulator (MI), especially when $J_{\rm{H}}/U$ is large.

\begin{figure}[t!]
\centering\includegraphics[
width=85mm 
]{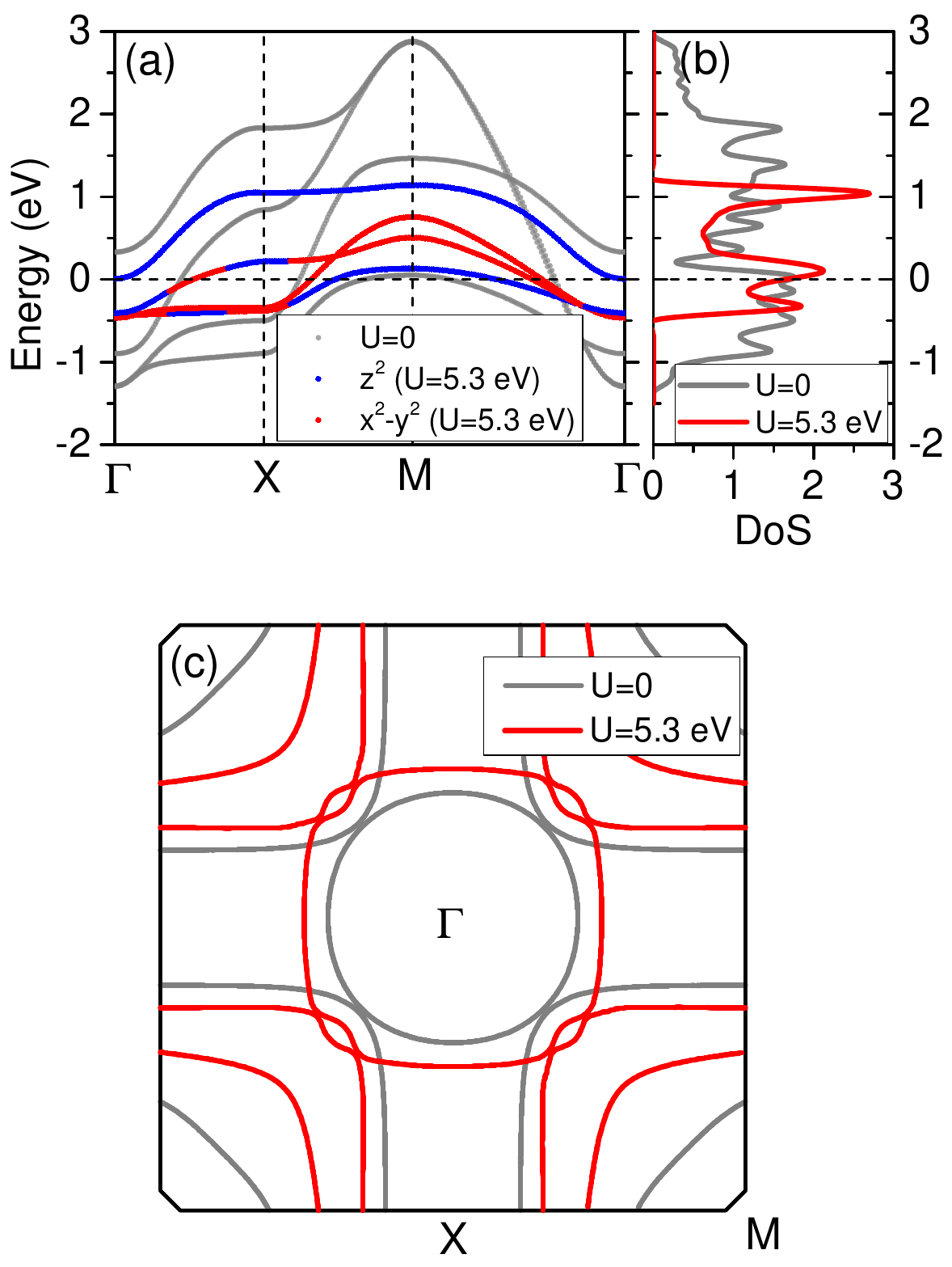}
\caption{(Color online) (a): Bandstructure at $U=5.3$ eV compared to the one at $U=0$,
indicating strong orbital-selective band renormalization.
(b): The corresponding electron density of states (DoS) showing a large renormalization on the bandwidth $W$.
(c): Comparison of the Fermi surface at $U=5.3$ eV and $U=0$.
}
\label{fig:3}
\end{figure}

We next consider the effects of orbital-selective Mott correlations to the electronic structure.
Fig.~\ref{fig:3}(a) shows the bands along high-symmetry directions of the Brillouin zone at $U=5.3$ eV
compared to those at $U=0$. Close to the OSMP, bands with the $x^2-y^2$ and $z^2$ bonding orbital characters
are strongly renormalized whereas the $z^2$ antibonding band, located topmost in energy,
only hardly shifts compared to the $U=0$ case. As shown in Fig.~\ref{fig:3}(b) and (c),
compared
to
the significant
renormalization in the
total bandwidth $W$ from about $4$ eV to about $1.5$ eV,
the Fermi surface only changes moderately.
While the relatively small
 inner hole pocket centered at the M point exhibits a sizable expansion,
 the outer hole and electron pockets
 centered around the
 $\Gamma$ point only slightly shrink and expand, respectively.

\begin{figure}[t!]
\centering\includegraphics[
width=85mm 
]{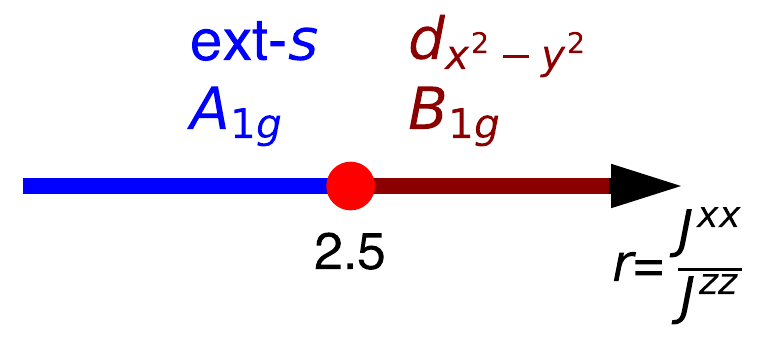}
\caption{(Color online) Evolution of the leading
superconducting
pairing symmetry,
from an extended $s$-wave $A^{1g}$ to a $d$-wave $B^{1g}$
with increasing $J^{xx}$. Here we take $J^{zz}=0.025W_0$, and vary the ratio $r=J^{xx}/J^{zz}$,
where $W_0$ is the total bandwidth of the multiorbital Hubbard model at $U=0$.
}
\label{fig:4}
\end{figure}

{\it An effective multiorbital $t$-$J$ model for superconductivity.~}
The above slave-spin results set the stage to build a low-energy effective model in understanding superconductivity of the system,
which can be done by performing a $t/U$ perturbation expansion when $U$ is sizeable. However, the resulting form of the effective theory
depends on the low-energy manifold the perturbed Hamiltonian is projected to.
For example, when projecting to the $S=1/2$ low-spin sector,
one ends up with an effective model that includes
 interactions between the total spin and isospin operators,
 which takes the form of the Kugel-Khomskii model~\cite{KK}.
 On the other hand, one obtains
three-orbital Heisenberg
couplings for the interacting part of the Hamiltonian when projecting to the high-spin sector.
Which model is pertinent to the low-energy physics depends on the strength of the interaction.

For La$_3$Ni$_2$O$_7$, $U$ is estimated to be within $4$ to $6$ eV~\cite{SunWang_Nature_2023, Werner_arXiv_2023}.
According to the phase diagram in Fig.~\ref{fig:1}(b), this suggests
that the
system is in the regime with strong orbital selectivity near the OSMP
(as highlighted in Fig.~\ref{fig:1}(b)).
Therefore, we construct the effective model by starting from the $S=3/2$ high-spin ground state,
and taking into account effects of the $S=1/2$ low-lying excitations
 by projecting out doubly occupied states.
As a result, it takes the form of a multiorbital $t$-$J$ model where the Hamiltonian reads
\begin{eqnarray}
\label{Eq:Ham_tJ}
H_{\rm{eff}} && = \frac{1}{2}\sum_{{ij}\alpha\beta\sigma} \sqrt{Z_{\alpha}Z_{\beta}} t^{\alpha\beta}_{ij}
 f^\dagger_{i\alpha\sigma} f_{j\beta\sigma} + \sum_{i\alpha\sigma} (\epsilon^\prime_\alpha-\mu)
 f^\dagger_{i\alpha\sigma} f_{i\alpha\sigma} \nonumber\\
&& + {\sum_{ij\alpha\beta}}^\prime J^{\alpha\beta}_{ij} \mathbf{S}_{i\alpha} \cdot \mathbf{S}_{j\beta}.
\end{eqnarray}
Here we have employed
the
slave-spin method to renormalize the kinetic part of the Hamiltonian: $Z_{\alpha}$ is the quasiparticle spectral weight of the $\alpha$-th MO, and $\epsilon^\prime_\alpha$ refers to the renormalized energy level of orbital $\alpha$. This is a generalization of the slave-boson theory~\cite{Lee_RMP_2006} to the finite $U$ case.
 $\mathbf{S}_{i\alpha}=\frac{1}{2} f^\dagger_{i\alpha s} \boldsymbol{\sigma}_{ss^\prime} f_{i\alpha s^\prime}$ denotes the spin density operator at site $i$ in the $\alpha$ orbital. Considering the high-spin state, the summation ${\sum}^\prime$ runs over the two $x^2-y^2$ and the one $z^2$ bonding MOs. $J^{\alpha\beta}_{ij}$ refers to the orbital dependent exchange coupling which can be determined from the second-order $t/U$ perturbation expansion~\cite{SM}. Here we neglect the inter-layer couplings between $x^2-y^2$ orbitals and second-nearest and further neighboring interactions because of their small hopping amplitudes, and only consider the in-plane nearest neighboring exchange interactions $J^{xx}_1$, $J^{zz}_1$, and $J^{xz}_1$. We find $J^{xz}_1$
 to be
 negligibly small, and $J^{xx}_1$, $J^{zz}_1$ are antiferromagnetic. In the following,
 we take the convention $J^{\alpha\beta}$ for $J^{\alpha\beta}_1$.
 To explore how the superconducting pairing evolves with orbital-selective correlations,
 we take $J^{zz}=0.025W_0$
 (with $W_0$ referring to the bandwidth at $U=0$),
and tune the ratio $r=J^{xx}/J^{zz}$.

{\it Superconducting pairing symmetry.~}
The superconducting pairing in the model of Eqn.~\eqref{Eq:Ham_tJ} is studied by a Bogoliubov
mean-field decomposition of the exchange interactions in the intra-orbital spin singlet sector:
\begin{eqnarray}
\label{Eq:Pairing}
J^{\alpha\alpha}_{\delta} (\mathbf{S}_{i\alpha} \cdot \mathbf{S}_{i+\delta\alpha}-\frac{1}{4} n_{i\alpha} n_{i+\delta\alpha}) ~~~~~~~~~~ \nonumber
\\
\approx -\frac{J^{\alpha\alpha}_{\delta}}{2} (\hat{\Delta}^\dagger_{\delta\alpha} \Delta_{\delta\alpha} + \rm{h.c.} -|\Delta_{\delta\alpha}|^2),
\end{eqnarray}
where $\hat{\Delta}_{\delta\alpha}=f_{i\alpha\downarrow}f_{i+\delta\alpha\uparrow}-f_{i\alpha\uparrow}f_{i+\delta\alpha\downarrow}$,
and the gap function $\Delta_{\delta\alpha}=\langle \hat{\Delta}_{\delta\alpha} \rangle$. La$_3$Ni$_2$O$_7$
under pressure has an orthorhomic structure. But the difference between the $a$ and $b$ lattice constants is small
(about $1\%$)~\cite{SunWang_Nature_2023}.
As such, for convenience, we examine the symmetry of the gap functions by studying how they transform
under the tetragonal $D_{4h}$ group.
The result is summarized in the SM (Tab.~S2)~\cite{SM}.
One sees that the multiorbital nature leads to six different pairing channels.
We then perform a
self-consistent calculation to determine the leading pairing channel~\cite{Yu_NC_2013}.
As shown in Fig.~\ref{fig:4}, the leading pairing channel changes from the extended $s$-wave $A^{1g}_{z^2}$ to $d$-wave $B^{1g}$
in both $x^2-y^2$ and $z^2$-bonding orbitals with increasing $J^{xx}$. In the $A^{1g}$ dominant regime,
the pairing is also strongly orbital-selective, with the leading channel associated with the $z^2$ bonding orbital.
Besides the larger pairing amplitude stabilized by $J^{zz}$, this pairing channel is also favored by causing a full superconducting gap
along the inner hole pocket centered at the M point. However, nodes along the outer hole pocket cannot be avoided by either pairing channel.
To avoid nodes, it is possible that a pairing
function
 with mixed $s$- and $d$-wave characters, such as the time-reversal breaking $s$+i$d$~\cite{Yu_NC_2013}
 or $s$+$d$ (given the orthorhmic lattice symmetry of the compound)~\cite{Hu_PRB_2018},
 is stabilized in the regime
 where $A^{1g}$ and $B^{1g}$ pairing channels are in
 competition.

{\it Discussions and conclusions.~} Several remarks are in order.
First, as shown in Fig.~\ref{fig:2}(a), near and inside the OSMP the $z^2$ bonding state is also very close to Mott localization. This makes the system to be in proximity to a multiorbital MI, which naturally explains the substantially suppressed Drude weight as observed in a recent optical conductivity measurement~\cite{Wen_arXiv_2023}. The strong orbital selectivity between the $z^2$ and $x^2-y^2$ orbitals accounts for the observed two-component contribution to the Drude weight~\cite{Wen_arXiv_2023}. Second, applying a pressure corresponding to increasing the hopping amplitudes, or equivalently, reducing the $U/t$ ratio in our model. This increases
the
itinerancy of electrons, and causes bandwidth tuning of the superconductivity. But in multiorbital systems, the effects of reducing the $U/t$ ratio
has additional effects. It can trigger a high-spin to low-spin crossover, as shown in the present work.
This activates the orbital degree of freedom, which makes the exchange couplings orbital dependent and may lead to strong competition
of fluctuations in the
antiferromagnetic spin
and orbital
channels
as reflected in the complicated temperature evolution of the magnetic susceptibility in La$_3$Ni$_2$O$_7$ at ambient pressure~\cite{Wu_PRB_2001,Liu_SC_2022}.
Moreover, reducing the $U/t$ ratio also causes redistribution of the electrons among the orbitals,
leading to an effect similar to either hole or electron doping a MI in each orbital.
This effect resembles a multiorbital version of the physics in doping the cuprates,
which is known to favor superconductivity.

In conclusion, we have studied electron correlation effects in a bilayer two-orbital Hubbard model for La$_3$Ni$_2$O$_7$ in the MO basis,
and
found a strong orbital selectivity when the interaction strength is moderate.
Further increasing the interaction, an OSMP is stabilized. The OSMP is close to an $S=3/2$ high-spin state, in which the $x^2-y^2$ and $z^2$ bonding orbitals are all very close to half-filling. In light of these results, we obtain an effective multiorbital $t$-$J$ model for superconductivity of the system
in the crossover regime towards the high-spin configuration. We show that
 the system exhibits orbital-selective pairing and the leading superconducting pairing channel evolves
  from the extended $s$-wave $A^{1g}$ to $d$-wave $B^{1g}$
 when the intra-orbital nearest-neighbor exchange coupling $J^{xx}$ is increased.
 Our work paves the way for systematically describing the
 pressure-induced high-$T_c$ superconductivity
 of La$_3$Ni$_2$O$_7$.

\begin{acknowledgments}
We
acknowledge Harold Hwang, Emilian M. Nica, Chandra Varma, Meng Wang, Hai-Hu Wen, and Weiqiang Yu
 for useful discussions. This work has in part been supported by
the National Science Foundation of China (Grants 12334008 and 12174441).
Work at Rice was primarily supported
by the U.S. Department of Energy, Office of Science, Basic Energy Sciences,
under Award No. DE-SC0018197, and by
the Robert A.\ Welch Foundation Grant No.\ C-1411.
Q.S. acknowledges the hospitality of the Aspen Center for Physics,
which is supported by NSF grant No. PHY-2210452,
during the workshop ``New directions on strange metals in correlated systems".

\end{acknowledgments}



 \clearpage
 \setcounter{figure}{0}
 \setcounter{table}{0}
\makeatletter
\renewcommand{\thefigure}{S\@arabic\c@figure}
\renewcommand{\thetable}{S\@arabic\c@table}
 \onecolumngrid

\section*{SUPPLEMENTAL MATERIAL -- Electron correlations and superconductivity in La$_3$Ni$_2$O$_7$ under pressure tuning}

\subsection{Details on the tight-binding model}


To include the realistic band structure at low energies into our tight-binding modeling,
we have first carried out band structure calculations for La$_3$Ni$_2$O$_7$
within the framework of density functional theory (DFT).
We have used the plane wave basis set as implemented in the Vienna Ab initio Simulation Package (VASP)
code \cite{vasp_website}. Projector augmented-wave potentials and Perdew-Burke-Ernzerhof exchange-correlation functional
were used in the calculations.
We consider the experimental lattice parameters ($a=5.289\;$\AA, $b=5.218\;$\AA, $c=19.734\;$\AA)~\cite{SunWang_Nature_2023}
in the simulations. Since the difference between $a$ and $b$ is only about 1.3$\%$, we use their average value ($a=b=5.2535\;$\AA)
in the calculations. Though this procedure has modified the space group from $Fmmm$ to $I4/mmm$, it should have little effect on the electronic structure.

\begin{figure}[th!]
\centering\includegraphics[
width=120mm 
]{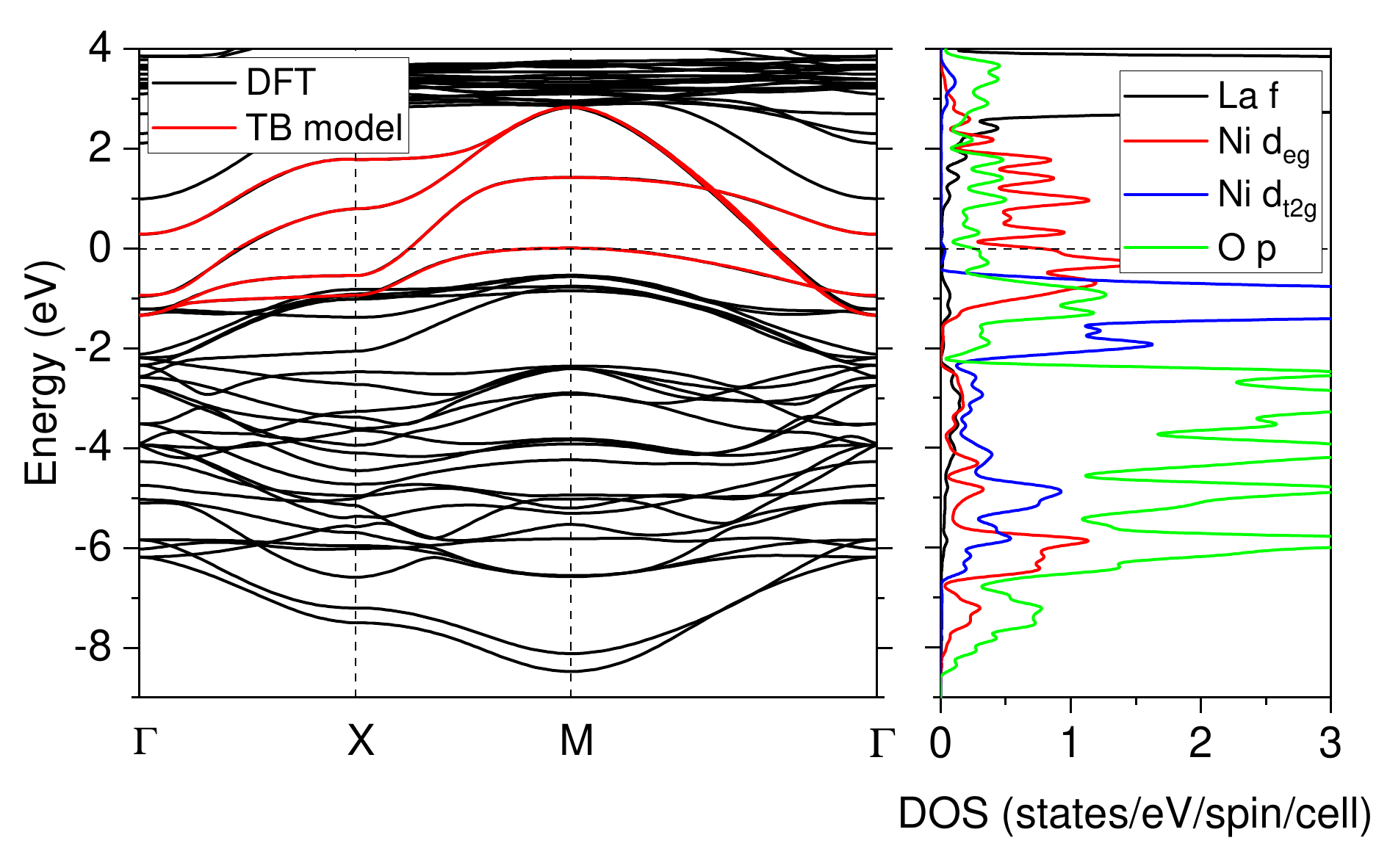}
\caption{(Color online) (a): Band structure of La$_3$Ni$_2$O$_7$ along high-symmetry directions of the BZ calculated
from DFT (black lines) and
from
the bilayer two-orbital tight-binding model (red lines).
(b): The corresponding projected density of states (DOS) calculated from DFT.
}
\label{fig:S1}
\end{figure}

As shown in Fig.~\ref{fig:S1}, the bands near the Fermi energy have mainly the Ni $3d$ $e_g$ orbital character.
The bands associated with
the
$t_{2g}$ orbitals are at least $1$ eV below the Fermi level. The oxygen $2p$ bands are dominant at about $4$ eV below the Fermi energy.
Therefore, the relevant orbitals within a $2$ eV energy window about the Fermi energy are the Ni $e_{g}$ ones.
We hence fit the Wannierized bands with a bilayer $2$-orbital tight-banding Hamiltonian including these $e_g$ orbitals
in the Brillouin zone (BZ) corresponding to the two Ni (in the top and bottom layers) unit cell.
At this step we have used projected Wannier functions; the procedure of disentanglement was performed
with the maximally-localized Wannier functions scheme as implemented in
the Wannier90 code \cite{Pizzi_JPCM_2020}. The tight-binding parameters from the fitting are summarized in Tab. S1.

The band structure of the bilayer 2-orbital tight-binding model compared to the DFT results is shown in Fig.~\ref{fig:S1}. The tight-binding model reproduces a similar band structure of DFT in the energy window
of interest, from $-1.5$ eV to $2.5$ eV.

\begin{table}[htbp]
		\centering
		\begin{tabular}{ccccccc}
			\hline\hline
           orbital & $\epsilon$ & $t_1$ & $t_2$ & $t_3$ & $t_{0z}$ & $t_{1z}$ \\
           \hline
           $xx$ & $10.8735$ & $-0.4899$ & $0.0670$ & $-0.0600$ & $0.0022$ & $0.0003$ \\
           $zz$ & $10.5142$ & $-0.1159$ & $-0.0109$ & $-0.0187$ & $-0.6363$ & $0.0231$ \\
           $xz$ &  & $\mp0.2420$ & $0$ & $\mp0.0333$ & $0$ & $\pm0.0382$ \\
           \hline\hline
		\end{tabular}
\caption{Tight-binding parameters of the bilyaer two-orbital model for La$_3$Ni$_2$O$_7$.
$xx$ ($zz$) denotes
the
 intra-orbital hopping between
the
$x^2-y^2$ ($z^2$) orbitals, whereas $xz$ denotes
the
 inter-orbital hopping between
 the $x^2-y^2$ and $z^2$ orbitals.
 $t_{n(z)}$ refers to intra- (inter-)layer hopping between the $n$-th neighboring sites.
 Finally,
 $\pm$ means
 that
  the hopping parameter is positive along the $x$ direction but negative along the $y$ direction.
 All units are in eV.}
\end{table}

\subsection{The interaction Hamiltonian in the bonding molecular orbital basis and the low-spin to high-spin crossover}
In the main text, we have performed a transformation from the atomic orbital
basis
 to the
 molecular orbital (MO) basis,
\begin{eqnarray}
 d_{i\alpha\sigma}^{b(a)} = \frac{1}{\sqrt{2}} (d_{i\alpha\sigma} \pm d_{i+\delta_{0z}\alpha\sigma}),
\end{eqnarray}
where the index $b(a)$ corresponds to the bonding (antibonding) MO, and $i$ ($i+\delta_{0z}$)
on the right hand side refers to a site in the top (bottom) layer. Here we define MOs for both
the
 $z^2$ and $x^2-y^2$ orbitals for convenience. But as listed in Tab. S1, the inter-layer hopping amplitude between
 the $x^2-y^2$ orbitals is much smaller than
 both its intra-layer counterpart and the inter-layer hopping between the $z^2$ orbitals.
 We therefore expect that the $x^2-y^2$ orbital is largely non-bonding.

In this MO basis, the interaction Hamiltonian is rewritten to
\begin{equation}
 H_{\rm{int}} = H^{b-b}_{\rm{int}} + H^{b-a}_{\rm{int}},
\end{equation}
where the boning-bonding and antibonding-antibonding
interactions are
\begin{eqnarray}
 	H^{b-b}_{int} &=& 
  \frac{U}{2} \sum_\alpha (n_{I\alpha\uparrow}^a n_{I\alpha\downarrow}^a + n_{I\alpha\uparrow}^b n_{I\alpha\downarrow}^b) + \frac{U'}{4} \sum_{\alpha\neq\beta,\sigma\neq\sigma'} (n_{I\alpha\sigma}^a n_{I\beta\sigma'}^a + n_{I\alpha\sigma}^b n_{I\beta\sigma'}^b)\nonumber\\
 	&+& \frac{U'-J}{2} \sum_{\alpha>\beta,\sigma} (n_{I\alpha\sigma}^a n_{I\beta\sigma}^a + n_{I\alpha\sigma}^b n_{I\beta\sigma}^b) - \frac{J}{2} \sum_{\alpha\neq\beta} (d_{I\alpha\uparrow}^{a+} d_{I\beta\downarrow}^{a+} d_{I\beta\uparrow}^{a} d_{I\alpha\downarrow}^{a} + d_{I\alpha\uparrow}^{b+} d_{I\beta\downarrow}^{b+} d_{I\beta\uparrow}^{b} d_{I\alpha\downarrow}^{b})\nonumber\\ 	&-&\frac{J}{2} \sum_{\alpha\neq\beta} (d_{I\alpha\uparrow}^{b+} d_{I\alpha\downarrow}^{b+} d_{I\beta\uparrow}^{b} d_{I\beta\downarrow}^{b} + d_{I\alpha\uparrow}^{a+} d_{I\alpha\downarrow}^{a+} d_{I\beta\uparrow}^{a} d_{I\beta\downarrow}^{a}),
\end{eqnarray}
and the bonding-antibonding mixing
interaction is
\begin{eqnarray}
 	H^{b-a}_{int} &=&	\frac{U}{2}\sum_\alpha(n_{I\alpha\uparrow}^an_{I\alpha\downarrow}^b+n_{I\alpha\uparrow}^bn_{I\alpha\downarrow}^a)+\frac{U'}{4}\sum_{\alpha\neq\beta,\sigma\neq\sigma'}(n_{I\alpha\sigma}^an_{I\beta\sigma'}^b+n_{I\alpha\sigma}^bn_{I\beta\sigma'}^a)	+\frac{U'-J}{2}\sum_{\alpha>\beta,\sigma}(n_{I\alpha\sigma}^an_{I\beta\sigma}^b+n_{I\alpha\sigma}^bn_{I\beta\sigma}^a)\nonumber\\
 	&+&\frac{U}{2}\sum_\alpha(d_{I\alpha\uparrow}^{b+}d_{I\alpha\uparrow}^ad_{I\alpha\downarrow}^{b+}d_{I\alpha\downarrow}^a+d_{I\alpha\uparrow}^{b+}d_{I\alpha\uparrow}^ad_{I\alpha\downarrow}^{a+}d_{I\alpha\downarrow}^b+d_{I\alpha\uparrow}^{a+}d_{I\alpha\uparrow}^bd_{I\alpha\downarrow}^{b+}d_{I\alpha\downarrow}^a+d_{I\alpha\uparrow}^{a+}d_{I\alpha\uparrow}^bd_{I\alpha\downarrow}^{a+}d_{I\alpha\downarrow}^b)\nonumber\\
 	 	&+&\frac{U'}{4}\sum_{\alpha\neq\beta,\sigma\neq\sigma'}(d_{I\alpha\sigma}^{b+}d_{I\alpha\sigma}^ad_{I\beta\sigma'}^{b+}d_{I\beta\sigma'}^a+d_{I\alpha\sigma}^{b+}d_{I\alpha\sigma}^ad_{I\beta\sigma'}^{a+}d_{I\beta\sigma'}^b+d_{I\alpha\sigma}^{a+}d_{I\alpha\sigma}^bd_{I\beta\sigma'}^{b+}d_{I\beta\sigma'}^a+d_{I\alpha\sigma}^{a+}d_{I\alpha\sigma}^bd_{I\beta\sigma'}^{a+}d_{I\beta\sigma'}^b)\nonumber\\
 	 	 	 	&+&\frac{U'-J}{2}\sum_{\alpha>\beta,\sigma}(d_{I\alpha\sigma}^{b+}d_{I\alpha\sigma}^ad_{I\beta\sigma}^{b+}d_{I\beta\sigma}^a+d_{I\alpha\sigma}^{b+}d_{I\alpha\sigma}^ad_{I\beta\sigma}^{a+}d_{I\beta\sigma}^b+d_{I\alpha\sigma}^{a+}d_{I\alpha\sigma}^bd_{I\beta\sigma}^{b+}d_{I\beta\sigma}^a+d_{I\alpha\sigma}^{a+}d_{I\alpha\sigma}^bd_{I\beta\sigma}^{a+}d_{I\beta\sigma}^b)\nonumber\\
 	 	 	 	&-&\frac{J}{2}\sum_{\alpha\neq\beta}(d_{I\alpha\uparrow}^{a+}d_{I\beta\downarrow}^{a+}d_{I\beta\uparrow}^{b}d_{I\alpha\downarrow}^b+d_{I\alpha\uparrow}^{b+}d_{I\beta\downarrow}^{b+}d_{I\beta\uparrow}^{a}d_{I\alpha\downarrow}^a+d_{I\alpha\uparrow}^{b+}d_{I\beta\downarrow}^{a+}d_{I\beta\uparrow}^{b}d_{I\alpha\downarrow}^a\nonumber\\
 	 	 	 	&+&d_{I\alpha\uparrow}^{b+}d_{I\beta\downarrow}^{a+}d_{I\beta\uparrow}^{a}d_{I\alpha\downarrow}^b+d_{I\alpha\uparrow}^{a+}d_{I\beta\downarrow}^{b+}d_{I\beta\uparrow}^{a}d_{I\alpha\downarrow}^b+d_{I\alpha\uparrow}^{a+}d_{I\beta\downarrow}^{b+}d_{I\beta\uparrow}^{b}d_{I\alpha\downarrow}^a)\nonumber\\	 		 	 	&-&\frac{J}{2}\sum_{\alpha\neq\beta}(d_{I\alpha\uparrow}^{a+}d_{I\alpha\downarrow}^{a+}d_{I\beta\uparrow}^{b}d_{I\beta\downarrow}^b+d_{I\alpha\uparrow}^{b+}d_{I\alpha\downarrow}^{b+}d_{I\beta\uparrow}^{a}d_{I\beta\downarrow}^a+d_{I\alpha\uparrow}^{b+}d_{I\alpha\downarrow}^{a+}d_{I\beta\uparrow}^{b}d_{I\beta\downarrow}^a\nonumber\\ 	 	 	 	&+&d_{I\alpha\uparrow}^{b+}d_{I\alpha\downarrow}^{a+}d_{I\beta\uparrow}^{a}d_{I\beta\downarrow}^b+d_{I\alpha\uparrow}^{a+}d_{I\alpha\downarrow}^{b+}d_{I\beta\uparrow}^{a}d_{I\beta\downarrow}^b+d_{I\alpha\uparrow}^{a+}d_{I\alpha\downarrow}^{b+}d_{I\beta\uparrow}^{b}d_{I\beta\downarrow}^a).	 	 	 	\end{eqnarray}

\begin{figure}[th!]
\centering\includegraphics[
width=160mm, trim=80 550 60 40,clip 
]{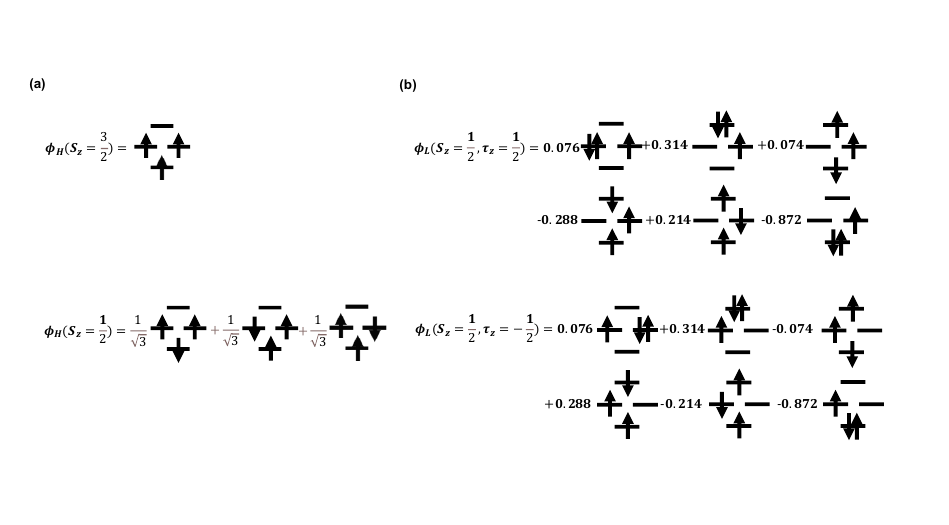}
\caption{(Color online) (a): High-spin $S=3/2$ ground-state configurations. (b): Low-spin $S=1/2$ ground-state configurations. Each ground state is four-fold degenerate, and the other degenerate configurations can be obtained by reversing the spin direction ($S^z\rightarrow -S^z$) from those presented here.
}
\label{fig:S2}
\end{figure}

Diagonalizing the interaction Hamiltonian along with the onsite potential term in the tight-binding Hamiltonian
written
 in the MO basis, we obtain two ground states in the atomic limit: an $S=1/2$ low-spin state and an $S=3/2$ high-spin state. These two ground states are illustrated in Fig.~\ref{fig:S2}. Each state is four-fold degenerate. In the $S=1/2$ low-spin state, the additional degeneracy is associated with the two degenerate $x^2-y^2$ orbitals. As one sees, the low-spin configuration is dominated by the Fock state that the bonding $z^2$ orbital is doubly occupied while the $x^2-y^2$ orbitals are quarter filled. We can then define an orbital isospin operator $\boldsymbol{\tau}$, with the Ising variable $\tau^z$ denoting the electron density difference in the $x^2-y^2$ orbitals between the top and bottom layers. The four degenerate low-spin configurations can be obtained from the one shown in Fig.~\ref{fig:S2} by appying total spin and orbital isospin reversal symmetry, respectively. On the other hand, the degenerate high-spin configurations can be labeled by the quantum number $S^z$ of the total spin.

\begin{figure}[th!]
\centering\includegraphics[
width=120mm 
]{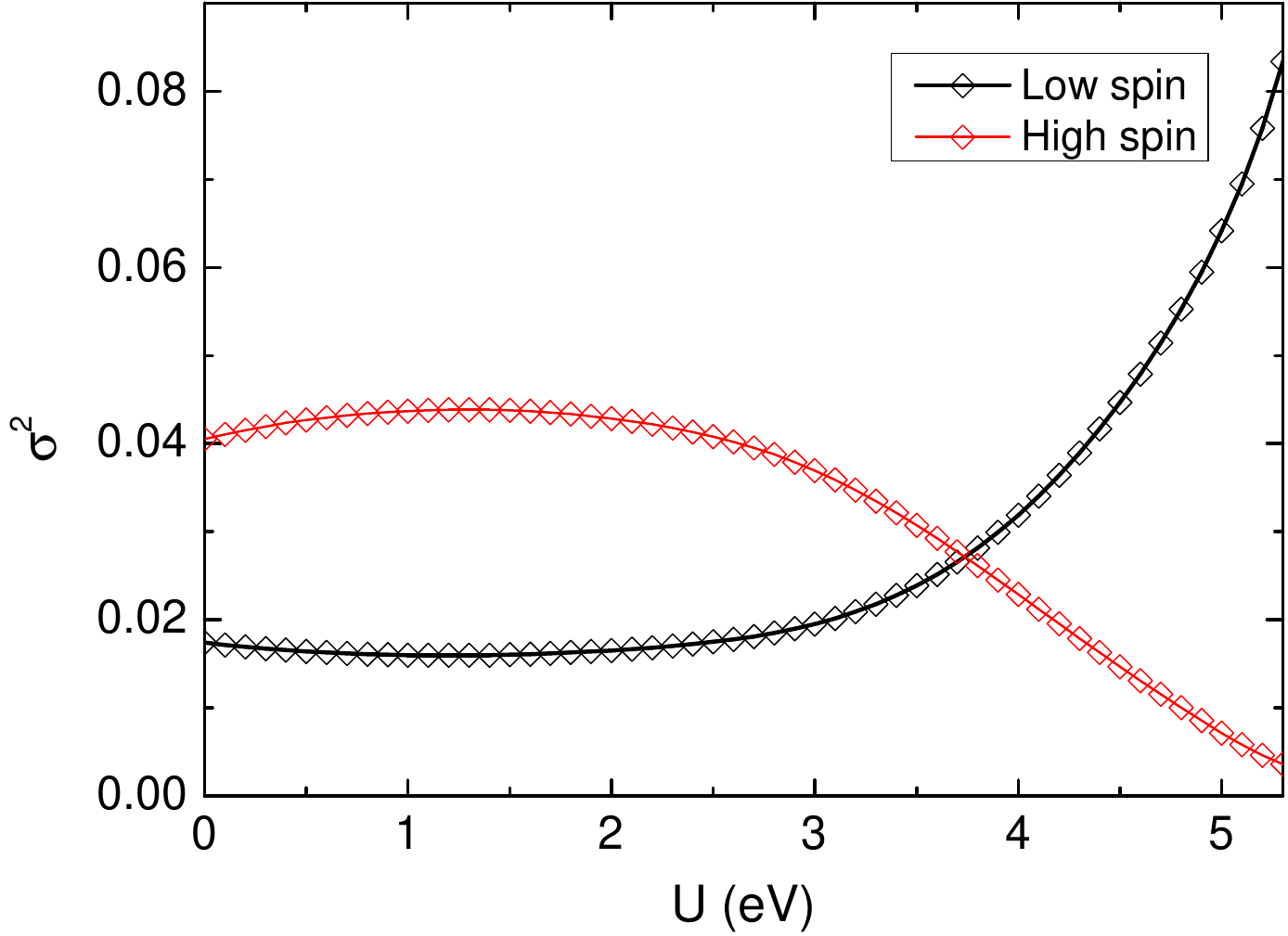}
\caption{(Color online) Variance of the electron density profile from the low-spin and high-spin states at $J_{\rm{H}}/U=0.25$. The low-spin to high-spin crossover is determined from the crossing point of the two variance curves.
}
\label{fig:S3}
\end{figure}

In the atomic limit, there is a transition from the low-spin to high-spin ground state by increasing $J_{\rm{H}}$, shown as the red line in the phase diagram of Fig.~1(b) in the main text. Taking into account the kinetic energy
will turn the transition to a crossover. To determine the crossover line in the phase diagram,
we define the variances of the electron density profile from that dominating the low- and high-spin states:
\begin{eqnarray}
 \sigma^2_{L} &&= \frac{1}{4}\left[(n_{z^2(b)}-1)^2 + (n_{z^2(a)})^2 + 2(n_{x^2-y^2}-1/4)^2)\right], \\
 \sigma^2_{H} &&= \frac{1}{4}\left[(n_{z^2(b)}-1/2)^2 + (n_{z^2(a)})^2 + 2(n_{x^2-y^2}-1/2)^2)\right].
\end{eqnarray}
The variances with $U$ at $J_{\rm{H}}/U=0.25$ is depicted in Fig.~\ref{fig:S3}. The low-spin to high-spin crossover is then determined by the criterion $\sigma^2_H=\sigma^2_L$, which gives the dashed line in the phase diagram of Fig.~1(b) in the main text.

\subsection{Details on the derivation of the effective model}

To construct the effective model, we start from the bilayer two-orbital Hubbard model in the MO basis, and rewrite the toal Hamiltonian into two parts, $H = H_0 + H_1$. We take the Hamiltonian in the atomic limit (including the interaction and onsite potential terms) as the unperturbed Hamiltonian $H_0$, and treat the hopping terms as perturbations ($H_1$). The effective low-energy model can be obtained via a canonical transformation,
	\begin{equation}
		H_{\rm{eff}} = e^{i\mathcal{S}}He^{-i\mathcal{S}}=H_0+[i\mathcal{S},H_0]+H_1+[i\mathcal{S},H_1] +\frac{1}{2}[i\mathcal{S},[i\mathcal{S},H_0]]+\dots
	\end{equation}
The unitary operator $\mathcal{S}$ can be determined by requiring the first-order contribution in $H_1$ to be $0$, and the effective Hamiltonian is derived from the second-order perturbation,
\begin{equation}
 H_{\rm{eff}} \approx \frac{1}{2}[i\mathcal{S},H_1].
\end{equation}

We should then project the effective Hamiltonian onto the low-energy subspace, which is the high-spin $S=3/2$ state in the strong-coupling limit $U/t\rightarrow\infty$. In this way, the effective model takes the form of
an $S=3/2$ Heisenberg model
	\begin{equation}\label{Eq:HeffH}
					H_{\rm{eff}} = \sum_{ij} J_{ij}\vec{S}_i\cdot \vec{S}_j, 
	\end{equation}
where $\vec{S}_i$ is the $S=3/2$ spin operator at the $i$-th unit cell.
While this model should be well applied in the limit $U/t\rightarrow\infty$,
calculations suggest the La$_3$Ni$_2$O$_7$ is close to the low-spin to high-spin crossover, where
the
effects of low-lying low-spin excitations are non-negligible. In practice,
 it would be difficult to project the effective Hamiltonian to a sector including both the high- and low-spin configurations
 given the incompatible quantum numbers of these two states.
 Here we adopt an alternative way. We consider a
 model with three-orbital Heisenberg
 couplings
	\begin{equation}\label{Eq:HeffM}
H_{\rm{eff}} = \sum_{ij,\alpha\beta} J^{\alpha\beta}_{ij} \vec{S}_{i\alpha} \cdot \vec{S}_{j\beta} - \tilde{J}_{\rm{H}} \sum_{i\alpha\beta} \vec{S}_{i\alpha} \cdot \vec{S}_{i\beta}, 
	\end{equation}
where $\vec{S}_{i\alpha}$ is an $S=1/2$ spin operator in orbital $\alpha$,
and $\alpha$ runs over the two $x^2-y^2$ and the bonding $z^2$ orbitals.
$\tilde{J}_{\rm{H}}$ is an effective Hund's coupling trying to align spin directions in all orbitals.
The
model in Eqn.~\eqref{Eq:HeffM} goes back to the one in Eqn.~\eqref{Eq:HeffH} in the limit
$\tilde{J}_{\rm{H}}\rightarrow\infty$. By reducing the value of $\tilde{J}_{\rm{H}}$,
the
effects from more low-spin configurations are taken into account.

To precisely determine the values of the model parameters $\tilde{J}_{\rm{H}}$ and $J^{\alpha\beta}_{ij}$
in Eqn.~\eqref{Eq:HeffM} requires accurate knowledge of $U$ and $J_{\rm{H}}$ in the original multiorbital Hubbard model.
We note that
the purpose of the present work is to capture the key features of superconductivity over a wide physical regime of model parameters.
As such, we further simplify the model in Eqn.~\eqref{Eq:HeffM} by taking $\tilde{J}_{\rm{H}}=0$. This makes
the spin of
each orbital to be
independent,
and
maximally includes effects from low-spin states. Following this way, we can project the effective Hamiltonian
to each orbital subspace independently and determine the orbital dependent effective exchange couplings.
For the in-plane nearest neighbor pair of sites, we find $J^{xx}_1=158.6$ meV, $J^{zz}_1=45.8$ meV,
and $J^{zx}_1=-1.1$ meV for $U=6$ eV and $J_{\rm{H}}/U=0.25$.
The
exchange couplings for a pair of sites along other directions are less than $1$ meV in magnitude and are hence neglected.
Note that for nearest neighbor pairs, the intra-orbital exchange couplings are both antiferromagnetic,
whereas the inter-orbital one is very weakly ferromagnetic. This implies a strong competition between antiferromagnetic
and ferromagnetic inter-orbital processes. Varying $U$ and $J_{\rm{H}}/U$ can significantly modify the effective exchange couplings.
In the calculation for superconductivity, we take $J^{zz}_1\approx 0.025W_0$ and leave $J^{xx}_1$ as a free parameter.
 Here $W_0\sim 4$ eV, is the bare bandwidth at $U=0$.

 \subsection{Symmetry classification of superconducting pairing functions}

For reasons given in the main text,
 we classify the symmetry of the superconducting gap functions by considering
how they transform
under the tetragonal $D_{4h}$ group. The corresponding result is given in
Tab.~\ref{Tab:S2}.

\begin{table}[t!]
		\centering
		\begin{tabular}{cccc}
			\hline\hline
			Orbital & $\Delta(\mathbf{r})$ & $\Delta(\mathbf{k})$ & Symmetry \\
			\hline
            $x^2-y^2$ & $(\Delta_{\hat{x}} + \Delta_{\hat{y}})\eta_0$ & $s_{x^2+y^2} \eta_0$ & $A^{1g}$ \\
             & $(\Delta_{\hat{x}} - \Delta_{\hat{y}})\eta_0$ & $d_{x^2-y^2} \eta_0$ & $B^{1g}$ \\
             & $(\Delta_{\hat{x}} + \Delta_{\hat{y}})\eta_3$ & $s_{x^2+y^2} \eta_3$ & $A^{2u}$ \\
             & $(\Delta_{\hat{x}} - \Delta_{\hat{y}})\eta_3$ & $d_{x^2-y^2} \eta_3$ & $B^{2u}$ \\
					\hline	
            $z^2$ & $\Delta_{\hat{x}} + \Delta_{\hat{y}}$ & $s_{x^2+y^2}$ & $A^{1g}$ \\
             & $\Delta_{\hat{x}} - \Delta_{\hat{y}}$ & $d_{x^2-y^2}$ & $B^{1g}$ \\
            \hline\hline
		\end{tabular}
\caption{Superconducting pairing symmetry of the effective multiorbital $t$-$J$ model. $\Delta(\mathbf{r})$ and $\Delta(\mathbf{k})$ refer to gap functions in real and momentum space, respectively. Their symmetry is characterized by the corresponding irreducible representation of the tetragonal $D_{4h}$ group. $\eta_0$ and $\eta_3$ refer to the $2\times2$ unit marix and $z$-component of the Pauli matrix, respectively. $\hat{x}(\hat{y})$ refers to unit vector along the $x(y)$ direction.}\label{Tab:S2}
\end{table}


\begin{thebibliography}{99}
\bibitem{Kamihara_JACS_2008} Y. Kamihara, T. Watanabe, M. Hirano, and H. Hosono, ``Iron-Based Layered Superconductor La[O$_{1-x}$F$_x$]FeAs ($x = 0.05$-$0.12$) with $T_{c} = 26$ K'', J. Am. Chem. Soc. {\bf 130}, 3296 (2008).

\bibitem{Johnston_2010}
D. C. Johnston, ``The puzzle of high temperature superconductivity in layered iron pnictides and
chalcogenides", Adv. Phys. {\bf 59}, 803 (2010).

\bibitem{Si-Hussey_2023}
Q. Si and N. E. Hussey, ``Iron-based superconductors: Teenage, complex, challenging",
Phys. Today {\bf 76}, 34 (2023).

\bibitem{SunWang_Nature_2023} H. Sun, M. Huo, X. Hu, J. Li, Z. Liu, Y. Han, L. Tang, Z. Mao, P. Yang, B. Wang, J. Cheng, D.-X. Yao, G.-M. Zhang, M. Wang, ``Signatures of superconductivity near 80 K in a nickelate under high pressure'',  Nature, (2023). \url{https://doi.org/10.1038/s41586-023-06408-7}

\bibitem{Cheng_arXiv_2023} J. Hou, P. T. Yang, Z. Y. Liu, J. Y. Li, P. F. Shan, L. Ma, G. Wang, N. N. Wang, H. Z. Guo, J. P. Sun, Y. Uwatoko, M. Wang, G. -M. Zhang, B. S. Wang, J. -G. Cheng, ``Emergence of high-temperature superconducting phase in the pressurized La$_3$Ni$_2$O$_7$ crystals'', arXiv:2307.09865 (2023).

\bibitem{Yuan_arXiv_2023} Yanan Zhang, Dajun Su, Yanen Huang, Hualei Sun, Mengwu Huo, Zhaoyang Shan, Kaixin Ye, Zihan Yang, Rui Li, Michael Smidman, Meng Wang, Lin Jiao, Huiqiu Yuan, ``High-temperature superconductivity with zero-resistance and strange metal behavior in La$_3$Ni$_2$O$_7$'', arXiv:2307.14819 (2023).

\bibitem{Li_Nature_2019} D. Li, K. Lee, B. Y. Wang, M. Osada, S. Crossley, H. R. Lee, Y. Cui, Y. Hikita, and H. Y. Hwang, ``Superconductivity in an infinite-layer nickelate'', Nature {\bf 572}, 624-627 (2019).

\bibitem{Yao_arXiv_2023} Zhihui Luo, Xunwu Hu, Meng Wang, Wei Wu and Dao-Xin Yao, ``Bilayer two-orbital model of La$_3$Ni$_2$O$_7$ under pressure", arXiv:2305.15564 (2023).

\bibitem{Hu_arXiv_2023} Yuhao Gu, Congcong Le, Zhesen Yang, Xianxin Wu and Jiangping Hu, ``Effective model and pairing tendency in bilayer Ni-based superconductor La$_3$Ni$_2$O$_7$", arXiv:2306.07275 (2023).

\bibitem{Werner_arXiv_2023} Viktor Christiansson, Francesco Petocchi and Philipp Werner, ``Correlated electronic structure of La$_3$Ni$_2$O$_7$ under pressure", arXiv:2306.07931 (2023).

\bibitem{WangQ_arXiv_2023} Qing-Geng Yang, Da Wang and Qiang-Hua Wang, ``Possible $S_{\pm}$-wave superconductivity in La$_3$Ni$_2$O$_7$", arXiv:2306.03706 (2023).

\bibitem{Lechermann_arXiv_2023} Frank Lechermann, Jannik Gondolf, Steffen B\"{o}tzel, and Ilya M. Eremin, ``Electronic correlations and superconducting instability in La$_3$Ni$_2$O$_7$ under high pressure'', arXiv:2306.05121 (2023).

\bibitem{ZhangG_arXiv_2023} Yang Shen, Mingpu Qin and Guang-Ming Zhang, ``Effective bi-layer model Hamiltonian and density-matrix renormalization group study for the high-T$_c$ superconductivity in La$_3$Ni$_2$O$_7$ under high pressure", arXiv:2306.07837 (2023).

\bibitem{Leonov_arXiv_2023} D. A. Shilenko and I. V. Leonov, ``Correlated electronic structure, orbital-selective behavior, and magnetic correlations in double-layer La$_3$Ni$_2$O$_7$ under pressure", arXiv:2306.14841 (2023).

\bibitem{Dagotto_arXiv_2023} Yang Zhang, Ling-Fang Lin, Adriana Moreo and Elbio Dagotto, ``Electronic structure, orbital-selective behavior, and magnetic tendencies in the bilayer nickelate superconductor La$_3$Ni$_2$O$_7$ under pressure", arXiv:2306.03231 (2023).

\bibitem{Kuroki_arXiv_2023} Hirofumi Sakakibara, Naoya Kitamine, Masayuki Ochi and Kazuhiko Kuroki, ``Possible high T$_c$ superconductivity in La$_3$Ni$_2$O$_7$ under high pressure through manifestation of a nearly-half-filled bilayer Hubbard model", arXiv:2306.06039 (2023).

\bibitem{Zhang_arXiv_2023} Aiqin Yang, Xiangru Tao, Yundi Quan and Peng Zhang, ``A first-principles investigation of the origin of superconductivity in Tl$Bi_2$", arXiv:2306.14365 (2023).

\bibitem{Lu_arXiv_2023} Xuejiao Chen, Peiheng Jiang, Jie Li, Zhicheng Zhong and Yi Lu, ``Critical charge and spin instabilities in superconducting La$_3$Ni$_2$O$_7$", arXiv:2307.07154 (2023).

\bibitem{YangF_arXiv_2023} Yu-Bo Liu, Jia-Wei Mei, Fei Ye, Wei-Qiang Chen and Fan Yang, ``The $s^{\pm}$-Wave Pairing and the Destructive Role of Apical-Oxygen Deficiencies in La$_3$Ni$_2$O$_7$ Under Pressure", arXiv:2307.10144 (2023).

\bibitem{Yang_arXiv_2023} Yingying Cao and Yi-feng Yang, ``Flat bands promoted by Hund's rule coupling in the candidate double-layer high-temperature superconductor La$_3$Ni$_2$O$_7$", arXiv:2307.06806 (2023).

\bibitem{Wang_arXiv_2023} Wei Wu, Zhihui Luo, Dao-Xin Yao and Meng Wang, ``Charge Transfer and Zhang-Rice Singlet Bands in the Nickelate Superconductor  La$_3$Ni$_2$O$_7$ under Pressure", arXiv:2307.05662 (2023).

\bibitem{WuYang_arXiv_2023} Chen Lu, Zhiming Pan, Fan Yang, and Congjun Wu, ``Interlayer coupling driven high-temperature superconductivity in La$_3$Ni$_2$O$_7$ under pressure'', arXiv:2307.14965 (2023).

\bibitem{Wen_arXiv_2023} Zhe Liu, Mengwu Huo, Jie Li, Qing Li, Yuecong Liu, Yaomin Dai, Xiaoxiang Zhou, Jiahao Hao, Yi Lu, Meng Wang, and Hai-Hu Wen, ``Electronic correlations and energy gap in the bilayer nickelate La$_3$Ni$_2$O$_7$'', arXiv:2307.02950 (2023).

\bibitem{SM} See Supplemental Material [http://link...] for details
about the tight-binding parameters, the interaction terms in the MO basis, the derivation of the multiorbital $t$-$J$ model, and the classification of the pairing functions, which include Refs.~\cite{SunWang_Nature_2023, vasp_website, Pizzi_JPCM_2020}.

\bibitem{Castellani_PRB_1978} C. Castellani, C. R. Natoli,  and J. Ranninger, ``Magnetic structure of V2O3 in the insulating phase'',
Phys. Rev. B \textbf{18}, 4945 (1978).

\bibitem{Yu_PRB_2012} R. Yu and Q. Si, ``$U(1)$ slave-spin theory and its application to Mott transition in a multiorbital model for iron pnictides'', Phys. Rev. B {\bf 86}, 085104 (2012).

\bibitem{Yu_PRB_2017} R. Yu and Q. Si, ``Orbital-selective Mott phase in multiorbital models for iron pnictides and chalcogenides'', Phys. Rev. B {\bf 96}, 125110 (2017).

\bibitem{KK} K. I. Kugel and D. I. Khomskii, Soviet Physics Uspekhi {\bf 25}, 231 (1982).

\bibitem{Lee_RMP_2006} Patrick A. Lee, Naoto Nagaosa, and Xiao-Gang Wen, ``Doping a Mott insulator: Physics of high-temperature superconductivity'', Rev. Mod. Phys. {\bf 78}, 17 (2006).

\bibitem{Yu_NC_2013} R. Yu, P. Goswami, Q. Si, P. Nikolic, and J.-X. Zhu, ``Superconductivity at the border of electron localization and itinerancy'', Nat Commun. {\bf 4}, 2783 (2013).

\bibitem{Hu_PRB_2018} H. Hu, R. Yu, E. M. Nica, J.-X. Zhu, and Q. Si, ``Orbital-selective superconductivity in the nematic phase of FeSe'', Phys. Rev. B {\bf 98}, 220503 (2018).

\bibitem{Wu_PRB_2001} Guoqing Wu, J. J. Neumeier, and M. F. Hundley, ``Magnetic susceptibility, heat capacity, and pressure dependence of the electrical resistivity of La$_3$Ni$_2$O$_7$ and La$_4$Ni$_3$O$_{10}$'', Phys. Rev. B {\bf 63}, 245120 (2001).

\bibitem{Liu_SC_2022} Z. Liu, H. Sun, M. Huo, X. Ma, Y. Ji, E. Yi, L. Li, H. Liu, J. Yu, Z. Zhang, Z. Chen, F. Liang, H. Dong, H. Guo, D. Zhong, B. Shen, S. Li, and M. Wang, ``Evidence for charge and spin density waves in single crystals of La$_3$Ni$_2$O$_7$ and La$_3$Ni$_2$O$_6$'', Sci. China Phys. Mech. Astr. {\bf 66}, 217411 (2023).

\bibitem{vasp_website} See VASP website [https://www.vasp.at/].

\bibitem{Pizzi_JPCM_2020} G. Pizzi {\it et al.}, ``Wannier90 as a community code: new features and applications'', J. Phys. Cond. Matt. {\bf 32}, 165902 (2020).
\end{thebibliography}
\end{document}